\documentclass[showpacs,preprintnumbers,amsmath,amssymb,APSl,prd,nofootinbib,superscriptaddress, twocolumn]{revtex4-2}

\usepackage{longtable}
\usepackage{bm}
\usepackage{amsfonts}
\usepackage{amsmath}
\usepackage{amssymb,epsf}
\usepackage{latexsym}
\usepackage{graphicx,epsfig}
\usepackage{amssymb}
\usepackage{float}
\usepackage{subfigure}
\usepackage{epstopdf}
\usepackage[colorlinks=true,citecolor=blue,linkcolor=red,urlcolor=black]{hyperref}
\usepackage{dcolumn}
\usepackage{wrapfig}
\usepackage{makeidx}
\usepackage{epsf}
\usepackage{color}
\usepackage{multirow}
\begin{document}
%%%%%%%%%%%%%%%%%%%%%%%%%%%%%%%%%%%%%%%%%%%%%%%%%%%%%%%%%%%%%%%%%%%%%%%%%%%%

\title{Junction conditions of Palatini $f\left(\mathcal R,T\right)$ gravity}

\author{Jo\~ao Lu\'is Rosa}
\email{joaoluis92@gmail.com}
\affiliation{Institute of Physics, University of Tartu, W. Ostwaldi 1, 50411 Tartu, Estonia}

\author{Diego Rubiera-Garcia} \email{drubiera@ucm.es}
\affiliation{Departamento de F\'isica Te\'orica and IPARCOS,
	Universidad Complutense de Madrid, E-28040 Madrid, Spain}

\date{\today}

%%%%%%%%%%%%%%%%%%%%%%%%%%%%%%%%%%%%%%%%%%%%%%%%%%%%%%%%%%%%%%%%%%%%%%%%%%%%
\begin{abstract} 
We work out the junction conditions for the Palatini $f(\mathcal{R},T)$ extension of General Relativity, where $f$ is an arbitrary function of the curvature scalar $\mathcal{R}$ of an independent connection, and of the trace $T$ of the stress-energy tensor of the matter fields. We find such conditions on the allowed discontinuities of several geometrical and matter quantities, some of which depart from their metric counterparts, and in turn extend their Palatini $f(\mathcal{R})$ versions via some new $T$-dependent terms. Moreover, we also identify some ``exceptional cases" of $f(\mathcal{R},T)$ Lagrangians such that some of these conditions can be discarded, thus allowing for further discontinuities in $\mathcal{R}$ and $T$ and, in contrast with other theories of gravity, they are shown to not give rise to extra components in the matter sector e.g. momentum fluxes and double gravitational layers. We discuss how these junction conditions, together with the non-conservation of the stress-energy tensor ascribed to these theories, may induce non-trivial changes in the shape of specific applications such as traversable thin-shell wormholes. 
\end{abstract}
%%%%%%%%%%%%%%%%%%%%%%%%%%%%%%%%%%%%%%%%%%%%%%%%%%%%%%%%%%%%%%%%%%%%%%%%%%%%

\maketitle

%%%%%%%%%%%%%%%%%%%%%%%%%%%%%%%%%%%%%%%%%%%%%%%%%%%%%%%%%%%%%%%%%%%%%%%%%%%%
\section{Introduction}\label{sec:intro}
%%%%%%%%%%%%%%%%%%%%%%%%%%%%%%%%%%%%%%%%%%%%%%%%%%%%%%%%%%%%%%%%%%%%%%%%%%%%

%%%%%%%%%%%%%%%%%%%%%%%%%%%%%%%%%%%%%%%%%%%%%%%%%%%%%%%%%%%%%%%%%%%%%%%%%%%%

In the search for gravitating solutions of physical interest, one is frequently challenged by the need to model a given system as two separate regions of space-time with similar or very different properties, matched at some separation hypersurface. This is the case, for instance, of stellar bodies \cite{GlenBook,Rosa:2020hex} and their gravitational collapse \cite{JoshiBook}, domain walls \cite{Vilenkin}, cosmic strings \cite{VilBook}, thin-shell wormholes \cite{Poisson:1995sv,Dias:2010uh,Rosa:2018jwp,Rosa:2021yym}, and so on. The mathematical consistence of such a modeling demands the functions characterizing both the gravitational and matter fields to satisfy a number of {\it junction} conditions at the matching hypersurface and across of it.  The specific shape of such conditions within the context of Einstein's General Theory of Relativity (GR) were originally obtained by Darmois \cite{Darmois} and much later rederived by Israel \cite{Israel}. 

Despite the success of GR in describing a wide variety of gravitational phenomena in astrophysical and cosmological scales, in the last few decades there has been a growing interest in considering alternative formulations of the gravitational interaction \cite{MG1,MG2,MG3,MG4} in order to address the supposed shortcomings of GR in certain regimes. Prominent among them are its ultraviolet completion on the strong-field limit \cite{QFT1,QFT2} and the related issue with space-time singularities \cite{Senovilla:2014gza}, the removal of dark sources into a consistent and observationally viable cosmology alternative to the $\Lambda$CDM paradigm \cite{Bull:2015stt}, or the existence of several types of black hole mimickers in the cosmic zoo \cite{Cardoso:2019rvt} to be searched for using multimessenger astronomy \cite{Addazi:2021xuf}. In order to extract specific predictions of every such theory within models of gravitating bodies, it is thus of interest to find the shape of its corresponding junction conditions, a task that must be done on a case-by-case basis, see e.g. \cite{Davis:2002gn,Deruelle:2007pt,Senovilla:2013vra,delaCruz-Dombriz:2014zaa,Reina:2015gxa,Vignolo:2018eco,Rosa:2021mln,Rosa:2021teg,Feng:2022rga} for some theories.

The enormous pool of modified theories of gravity at our disposal can be classified according to the underlying hypothesis of GR being modified. This includes, but it is not limited to, modifications of the functional form of the action, introduction of new (dynamical) degrees of freedom, violations of mathematical/physical principles (e.g. Lorentz invariance), or the inclusion/re-consideration of additional geometrical ingredients. The latter includes the so-called metric-affine (or Palatini) formulation of gravity, in which metric and affine connection are restored to their roles as independent entities \cite{Olmo:2011uz}. While for the Einstein-Hilbert action of GR this is harmless since the Palatini version is fully equivalent to the metric one (up to a projective mode \cite{Bejarano:2019zco}), this is not so for other functional dependences of the action on scalar objects. This is the case, for instance, for $f(\mathcal{R})$ gravity (with $\mathcal{R}$ the Ricci scalar of an independent connection), whose dynamics are completely different from their metric counterparts. This is also reflected on some crucial aspects of their corresponding junction equations \cite{Olmo:2020fri}, which in turn introduces large qualitative differences in the implementation of specific applications, like stellar surfaces or thin-shell wormholes \cite{Olmo:2020fri,Lobo:2020vqh}. 

The main aim of this work is to extend the range of applicability of such junction conditions to a generalization of the latter theories including an additional contribution in the trace $T$ of the stress-energy tensor. Such theories were first introduced in the metric formalism as $f(R,T)$ gravity \cite{Harko:2011kv}  (with $R$ the usual metric curvature scalar) and more recently considered in the Palatini one as $f(\mathcal{R},T)$ gravity \cite{Barrientos:2018cnx,Wu:2018idg}. Both formulations introduce an extra force supposedly induced by exotic fluids or quantum effects, but in any case leading to a non-geodesic motion on the strong-field regime. Since the exploration of such hypothetical effects is of interest for the sake of cosmological \cite{Jamil:2011ptc,Zaregonbadi:2016xna,Singh:2018xjv,Varshney:2020eun,Bhattacharjee:2020jsf,Gamonal:2020itt,Rosa:2021tei,Rosa:2021myu,Rosa:2022fhl,Goncalves:2021vci,Goncalves:2022ggq} and astrophysical \cite{Moraes:2015uxq,Moraes:2017mir,Deb:2017rhd,Moraes:2017mir,Das:2017rhi,Elizalde:2018frj,Maurya:2019hds,Pretel:2020oae,Pretel:2021kgl} applications, it is worth finding the corresponding junction conditions in Palatini $f(\mathcal{R},T)$ gravity. To this end we shall use both the geometrical and its physically equivalent scalar-tensor representation, supplying ourselves with the formalism of tensorial distributions. We find the same number of conditions as in the Palatini $f(\mathcal{R})$ case, though in two of them corrections via derivatives of the gravity function with respect to $T$ are present.  Furthermore we identify several classes of ``exceptional" $f(\mathcal{R},T)$ Lagrangians in which some of these conditions can be discarded, thus allowing for additional discontinuities in geometrical and/or matter quantities. As a specific application, we discuss the case of thin-shell wormholes sourced by electromagnetic (Maxwell) fields, finding that no deviations are to be expected as compared to their $f(\mathcal{R})$ counterparts, but that if the electromagnetic field is allowed to be non-linear with $T \neq 0$ then every non-trivial function $f(\mathcal{R},T)$ will lead to new dynamics. In both of these cases  the non-conservation of the stress-energy tensor should also lead to differences in the structure of such wormholes as compared to the $f(R)$ \cite{Lobo:2009ip,Pavlovic:2014gba,Bahamonde:2016ixz,Godani:2020vqe}, $f(R,T)$ \cite{Zubair:2016cde,Sahoo:2017ual,Mishra:2019jpa,Banerjee:2019wjj} and  $f(\mathcal{R})$ \cite{Lobo:2020vqh} cases.

This paper is organized as follows: in Sec. \ref{sec:theory} we introduce the $f(\mathcal{R},T)$ theory in both the geometrical and the scalar-tensor representation and obtain their respective equations of motion; in Sec. \ref{sec:junctions} we derive the junction conditions of the theory in both representations using the distribution formalism, including some ``exceptional cases" for which the set of junction conditions can be simplified; in Sec. \ref{sec:apps} we provide an application of the junction conditions for the case of a thin-shell wormhole; and in Sec. \ref{sec:concl} we depict our conclusions.

%%%%%%%%%%%%%%%%%%%%%%%%%%%%%%%%%%%%%%%%%%%%%%%%%%%%%%%%%%%%%%%%%%%%%%%%%%%%
\section{Theory and equations}\label{sec:theory}
%%%%%%%%%%%%%%%%%%%%%%%%%%%%%%%%%%%%%%%%%%%%%%%%%%%%%%%%%%%%%%%%%%%%%%%%%%%%

%%%%%%%%%%%%%%%%%%%%%%%%%%%%%%%%%%%%%%%%%%%%%%%%%%%%%%%%%%%%%%%%%%%%%%%%%%%%
\subsection{Geometrical representation}\label{sec:geo}
%%%%%%%%%%%%%%%%%%%%%%%%%%%%%%%%%%%%%%%%%%%%%%%%%%%%%%%%%%%%%%%%%%%%%%%%%%%%

We consider the action of Palatini $f(\mathcal R,T)$ gravity in the geometric representation of the theory, defined as
\begin{equation}\label{geo_action}
\mathcal{S}=\frac{1}{2\kappa^2}\int_\mathcal V d^4x \sqrt{-g}f(\mathcal R,T)+\int_\mathcal V d^4x \sqrt{-g} \mathcal{L}_m(g_{\mu\nu},\psi_m) \ ,
\end{equation}
where $\kappa^2=8\pi G/c^4$, with $G$ the gravitational constant and $c$ the speed of light, while $\mathcal V$ is the space-time manifold and  $g$ is the determinant of the metric $g_{ab}$ written in terms of a coordinate set $x^a$. The gravitational Lagrangian is any well-behaved function $f(\mathcal R,T)$ of the Ricci scalar $\mathcal R=g^{ab}\mathcal R_{ab}$ built from the Ricci tensor of an independent connection $\hat \Gamma$, and written in the usual form
\begin{equation}\label{def_palricci}
\mathcal R_{ab}=\partial_c\hat\Gamma^c_{ab}-\partial_b\hat\Gamma^c_{ac}+\hat\Gamma^c_{cd}\hat\Gamma^d_{ab}-\hat\Gamma^c_{ad}\hat\Gamma^d_{cb} \ ,
\end{equation}
while $T=g^{ab}T_{ab}$ is the trace of the stress-energy tensor $T_{ab}$. Finally, $\mathcal L_m$ is the matter Lagrangian of a set of fields $\psi_m$, and coupled to the metric $g_{ab}$ (but not to the connection $\hat \Gamma$). In the following, we shall consider a geometrized unit system for which $G=c=1$, and thus $\kappa^2=8\pi$.

The action in Eq. \eqref{geo_action} depends on two independent entities, namely, the space-time metric $g_{ab}$ and the affine connection $\hat\Gamma$, each of which is to be described by its own equation of motion. This way, taking a variation of Eq. \eqref{geo_action} with respect to the metric $g_{ab}$, one obtains the modified field equations as
\begin{equation}\label{geo_field}
f_\mathcal{R}\mathcal R_{ab}-\frac{1}{2}f(\mathcal R,T)g_{ab}=8\pi T_{ab}-f_T\left(T_{ab}+\Theta_{ab}\right) \ ,
\end{equation} 
where the subscripts $\mathcal R$ and $T$ denote partial derivatives of the function $f(\mathcal R,T)$ with respect to these variables, i.e., $f_\mathcal R\equiv\partial f/\partial\mathcal R$ and $f_T\equiv\partial f/\partial T$. Two contributions appear in the right-hand side of these equations: $T_{ab}$ is defined in terms of the variation of the matter Lagrangian $\mathcal L_m$ with respect to the metric $g_{ab}$ in the usual way as
\begin{equation}\label{def_tab}
T_{ab}=-\frac{2}{\sqrt{-g}}\frac{\delta\left(\sqrt{-g}\mathcal L_m\right)}{\delta g^{ab}}=-2\frac{\partial \mathcal{L}_m}{\delta g^{ab}}+g_{ab}\mathcal{L}_m \ ,
\end{equation}
while the tensor $\Theta_{ab}$ is defined in terms of the variation of $T_{ab}$ with respect to the metric $g_{ab}$, i.e., 
\begin{equation}\label{def_theta}
\Theta_{ab}=g^{cd}\frac{\delta T_{cd}}{\delta g^{ab}}=g_{ab}\mathcal{L}_m-2T_{ab}-2g^{cd}\frac{\partial^2 \mathcal{L}_m}{\partial g^{ab}g^{cd}} \ .
\end{equation}
It should be stressed that the right-hand side of  the field equations in Eq. (\ref{geo_field}) can be read off as an effective stress-energy tensor of the form
\begin{equation} \label{eq:taumunu}
\tau_{ab}=T_{ab}\left(1-\frac{f_T}{8\pi}\right) -\frac{f_T}{8\pi}\Theta_{ab} \ ,
\end{equation}
whose divergence reads  \cite{Barrientos:2018cnx,Wu:2018idg}
\begin{equation}
\nabla_{a}{\tau^a}_{b}=-\frac{f_T}{16\pi} \nabla_{b}T \ , \label{eq: non-con}
\end{equation}
which is, in general, non-vanishing. Therefore, in Palatini $f(\mathcal{R},T)$ gravity, the effective stress-energy tensor is not conserved unless the product $f_T \partial_b T$ vanishes, a similar result as in the metric formulation of these theories \cite{Harko:2011kv}. Moreover, this effective stress-energy tensor plays another role in the link between the curvature and the behaviour of the matter fields, since contracting in Eq. (\ref{geo_field}) with $g^{ab}$ provides the result
\begin{equation} \label{eq:curtra}
\mathcal{R}f_{\mathcal{R}}-2f(\mathcal{R},T)=\kappa^2 \tau \ ,
\end{equation}
For a general function $f(\mathcal{R},T)$ this is an algebraic equation allowing to write $\mathcal{R}=\mathcal{R}(T,\Theta)$, i.e., the Palatini curvature can be removed in favour of the trace of the two objects $T_{ab}$ and $\Theta_{ab}$.

In order to be able to compute an explicit form of the tensor $\Theta_{ab}$, it is necessary to set first an explicit form of the tensor $T_{ab}$ or, equivalently, of the matter Lagrangian $\mathcal{L}_m$. For instance, let us consider  a stress-energy tensor that describes an anisotropic fluid, which is given by the expression
\begin{equation}\label{def_anisotropic}
T_{ab}=\left(\rho+p_{\perp}\right)u_au_b+p_{\perp}g_{ab}+\left(p_r-p_{\perp}\right)v_av_b \ ,
\end{equation}
where $\rho$ is the energy density, $p_r$ is the radial pressure, $p_{\perp}$ is the tangential pressure, $u^a$ is the four-velocity vector, and $v^a$ is the radial four-vector. Under these conditions, the matter Lagrangian can be written in the form $\mathcal L_m=\frac{1}{3}\left(2 p_{\perp}+p_r\right)$ \cite{Deb:2018sgt}, and the tensor $\Theta_{ab}$ becomes
\begin{equation} \label{eq:thetadef}
\Theta_{ab}=-2 T_{ab}+\frac{1}{3}g_{ab}\left(2 p_{\perp}+p_r\right) \ .
\end{equation}

On the other hand, taking a variation of Eq. \eqref{geo_action} with respect to the independent connection $\hat\Gamma$ yields the equation of motion
\begin{equation}\label{geo_gammaeom}
\hat\nabla_c\left(\sqrt{-g}f_\mathcal{R}g^{ab}\right)=0 \ ,
\end{equation}
where $\hat\nabla_c$ denotes covariant derivatives written in terms of the connection $\hat\Gamma$. Since $\sqrt{-g}$ is a scalar density of weight 1, its covariant derivative vanishes, i.e., $\hat\nabla_c\sqrt{-g}=0$, and this factor can be eliminated from Eq. \eqref{geo_gammaeom}. Defining a new metric tensor $\hat g_{ab}=f_\mathcal{R}g_{ab}$, Eq. \eqref{geo_gammaeom} then reduces to $\hat\nabla_c \hat g_{ab}=0$. This implies that $\hat\Gamma$ is the Levi-Civita connection of the new metric $\hat g_{ab}$, i.e., one can write
\begin{equation}\label{def_hatgamma}
\hat\Gamma^a_{bc}=\frac{1}{2}\hat g^{ad}\left(\partial_b\hat g_{dc}+\partial_c\hat g_{bd}-\partial_d\hat g_{bc}\right) \ .
\end{equation}
As the two metrics $\hat g_{ab}$ and $g_{ab}$ are conformally related to each other with a conformal factor $f_\mathcal R$, this implies that their corresponding Ricci tensors $\mathcal R_{ab}$ and $R_{ab}$, although assumed to be independent \textit{a priori}, are in fact related to each other via 
\begin{equation}\label{def_riccirel}
\mathcal R_{ab}=R_{ab}-\frac{1}{f_\mathcal R}\left(\nabla_a\nabla_b+\frac{1}{2}g_{ab}\Box\right)f_\mathcal R+\frac{3}{2f_\mathcal R^2}\partial_af_\mathcal R\partial_bf_\mathcal R \ ,
\end{equation}
where $\Box=\nabla_a\nabla^a$ represents the d'Alembert operator. Equations \eqref{def_riccirel} and \eqref{geo_gammaeom} are equivalent, and thus we shall use Eq. \eqref{def_riccirel} from this point onwards due to its simpler structure and ease to use.

%%%%%%%%%%%%%%%%%%%%%%%%%%%%%%%%%%%%%%%%%%%%%%%%%%%%%%%%%%%%%%%%%%%%%%%%%%%%
\subsection{Scalar-tensor representation}\label{sec:scalar}
%%%%%%%%%%%%%%%%%%%%%%%%%%%%%%%%%%%%%%%%%%%%%%%%%%%%%%%%%%%%%%%%%%%%%%%%%%%%

It is often useful to recast Eq. \eqref{geo_action} in a dynamically equivalent scalar-tensor representation, which was proven useful in many  modified theories of gravity, particularly those featuring extra scalar degrees of freedom beyond those of GR. This can be done by introducing two auxiliary fields $\alpha$ and $\beta$ in the action as
\begin{eqnarray}
S&=&\frac{1}{2\kappa^2}\int_\Omega\sqrt{-g}\left[f(\alpha,\beta)+f_\alpha(\mathcal R-\alpha)+\right.\nonumber \\
&+&\left.f_\beta(T-\beta)\right]d^4x+\int_\Omega\sqrt{-g}\mathcal L_m d^4x \ ,\label{st_auxaction1}
\end{eqnarray}
where the subscripts $\alpha$ and $\beta$ denote partial derivatives of the function $f\left(\alpha,\beta\right)$ with respect to these fields, i.e., $f_\alpha\equiv\partial f/\partial \alpha$ and $f_\beta\equiv\partial f/\partial \beta$. Eq. \eqref{st_auxaction1} depends on four independent quantities, namely the metric $g_{ab}$, the connection $\hat\Gamma$, and the two fields $\alpha$ and $\beta$. Taking a variation with respect to $\alpha$ and $\beta$ yields two coupled equations of motion
\begin{eqnarray}
&&f_{\alpha\alpha}\left(\mathcal R-\alpha\right)+f_{\alpha\beta}\left(T-\beta\right)=0 \ , \label{st_eomalpha} \\
&&f_{\beta\alpha}\left(\mathcal R-\alpha\right)+f_{\beta\beta}\left(T-\beta\right)=0 \ . \label{st_eombeta}
\end{eqnarray}
The system of Eqs. \eqref{st_eomalpha} and \eqref{st_eombeta} can be rewritten in a matrix form $\mathcal M \textbf{x}=0$ as
\begin{equation}\label{st_matrixeq}
\mathcal M\textbf{x}=\begin{pmatrix}
f_{\alpha\alpha} & f_{\alpha\beta} \\
f_{\beta\alpha} & f_{\beta\beta}
\end{pmatrix}
\begin{pmatrix}
\mathcal R-\alpha \\
T-\beta
\end{pmatrix}
=0 \ .
\end{equation}
For any general function $f\left(\alpha,\beta\right)$ satisfying the Schwartz theorem, i.e., for which the high-order derivatives are commutative, that is, $f_{\alpha\beta}=f_{\beta\alpha}$, the solution of Eq. \eqref{st_matrixeq} will be unique if and only if the determinant of the matrix $\mathcal M$ is non-vanishing, i.e., $f_{\alpha\alpha}f_{\beta\beta}-f_{\alpha\beta}^2\neq0$. Whenever this condition is satisfied, the unique solution of Eq. \eqref{st_matrixeq} is $\alpha=\mathcal R$ and $\beta=T$. Inserting these solutions back into Eq. \eqref{st_auxaction1}, one recovers Eq. \eqref{geo_action}, thus proving that the two formalisms are equivalent. If, on the other hand, the determinant of $\mathcal M$ vanishes, the solution of Eq. \eqref{st_matrixeq} is not unique and the scalar-tensor representation can no longer be guaranteed to represent the same theory as the geometrical representation.

It is now useful to introduce the following definitions for the two scalar fields $\varphi$ and $\psi$ and a scalar interaction potential $V\left(\varphi,\psi\right)$ as
\begin{eqnarray}
\varphi&=&\frac{\partial f}{\partial \mathcal R}, \qquad \psi=\frac{\partial f}{\partial T}, \label{def_scalars} \\
V\left(\varphi,\psi\right)&=&-f\left(\alpha,\beta\right)+\alpha\varphi+\beta\psi \ . \label{def_potential}
\end{eqnarray}
Introducing the definitions of Eqs. \eqref{def_scalars} and \eqref{def_potential} into Eq. \eqref{st_auxaction1}, considering the unique solution $\alpha=\mathcal R$ and $\beta=T$, and using the trace of Eq. \eqref{def_riccirel} to eliminate $\mathcal R$ in terms of the Ricci scalar $R$ of the metric $g_{ab}$, one obtains the action that defines the scalar-tensor representation of the Palatini $f\left(\mathcal R,T\right)$ gravity as
\begin{eqnarray}
S&=&\frac{1}{2\kappa^2}\int_\Omega\sqrt{-g}\Big[\varphi R+\frac{3}{2\varphi}\partial_a\varphi\partial^a\varphi\label{st_action}\\
&+& \psi T-V(\varphi,\psi)\Big]d^4x+\int_\Omega\sqrt{-g}\mathcal L_m d^4x\nonumber \ .
\end{eqnarray}
Note that the term proportional to $\Box f_\mathcal R$ in the trace of Eq. \eqref{def_riccirel} does not contribute to the action since it can be rewritten as a boundary term which vanishes by definition. It is useful to note that, similarly to what happens in the Palatini approach to $f\left(\mathcal{R}\right)$ gravity, the scalar-tensor representation of the Palatini $f\left(\mathcal{R},T\right)$ gravity features a scalar field $\varphi$ analogous to the Brans-Dicke scalar field with a parameter $\omega_{BD}=-3/2$. The difference to the scalar-tensor representation of Palatini $f\left(\mathcal{R}\right)$ gravity is the existence of a second scalar field $\psi$ associated to the arbitrary dependence of the action in $T$, contributing to the action with an extra term. 

Note that the addition of the scalar field $\varphi$ effectively removes the dependence of Eq.  \eqref{st_action} in $\hat\Gamma$, which is now dependent only in $g_{ab}$ and the scalar fields $\varphi$ and $\psi$. Taking the variation of Eq. \eqref{st_action} with respect to the metric $g_{ab}$ yields the field equations
\begin{eqnarray}
&&\varphi G_{ab}+\frac{3}{2\varphi}\left(\nabla_a\varphi\nabla_b\varphi-\frac{1}{2}g_{ab}\nabla_a\varphi\nabla^a\varphi\right)+\frac{1}{2}g_{ab}V \label{st_field}\\
&&-\left(\nabla_a\nabla_b-g_{ab}\Box\right)\varphi=8\pi T_{ab}-\psi\left(T_{ab}+\Theta_{ab}+\frac{1}{2}g_{ab}T\right)\nonumber,
\end{eqnarray}
where we have introduced the Einstein's tensor $G_{ab}=R_{ab}-\frac{1}{2}Rg_{ab}$ and the tensors $T_{ab}$ and $\Theta_{ab}$ have been previously defined in Eqs. \eqref{def_tab} and \eqref{def_theta}, respectively. On the other hand, the variations with respect to the scalar fields $\varphi$ and $\psi$ yield the following equations of motion
\begin{eqnarray}
&&\Box\varphi-\frac{1}{2\varphi}\nabla^a\varphi\nabla_a\varphi-\frac{\varphi}{3}
\left(R-V_\varphi\right)=0 \ , \label{st_eomphi} \\
&&T=V_\psi \ , \label{st_eompsi}
\end{eqnarray}
where the subscripts $\varphi$ and $\psi$ denote partial derivatives of $V(\varphi,\psi)$ with respect to these fields, i.e., $V_\varphi\equiv \partial V/\partial\varphi$ and $V_\psi\equiv\partial V/\partial \psi$ respectively. From Eq. \eqref{st_eomphi}, one might suspect that the scalar field $\varphi$ is dynamical, due to the existence of a term proportional to $\Box\varphi$. However, similarly to what happens in Palatini $f(\mathcal{R})$ gravity, one can show this not to be true. Taking the trace of Eq. \eqref{st_field} and using the result to eliminate $R$ from Eq. \eqref{st_eomphi}, one verifies that the terms $\Box\varphi$ and $\nabla_a\varphi\nabla^a\varphi$ cancel out, leading to
\begin{equation}\label{st_eomphi2}
\frac{1}{3}\left(2V-\varphi V_\varphi\right)=\frac{8\pi}{3}T-\frac{\psi}{3}\left(T+\Theta\right),
\end{equation}
thus proving that the scalar field $\varphi$ is not dynamical.

%%%%%%%%%%%%%%%%%%%%%%%%%%%%%%%%%%%%%%%%%%%%%%%%%%%%%%%%%%%%%%%%%%%%%%%%%%%%
\section{Junction conditions}\label{sec:junctions}
%%%%%%%%%%%%%%%%%%%%%%%%%%%%%%%%%%%%%%%%%%%%%%%%%%%%%%%%%%%%%%%%%%%%%%%%%%%%

\subsection{Notation and assumptions}

Let us start our analysis of the junction conditions by specifying the notation and assumptions to be used. Let $\Sigma$ be a hypersurface that separates the whole spacetime $\mathcal V$ into two regions, $\mathcal V^+$ and $\mathcal V^-$. Consider that the metric $g_{ab}^+$, expressed in coordinates $x^a_+$, is the metric in the region $\mathcal V^+$, and the metric $g_{ab}^-$, expressed in coordinates $x^a_-$, is the metric in the region $\mathcal V^-$, with latin indexes running from $0$ to $3$. Assume that a set of coordinates $y^\alpha$ can be defined in both sides of $\Sigma$, with greek indexes excluding the index in the direction perpendicular to $\Sigma$. The projection vectors from the 4-dimensional regions $\mathcal V^\pm$ to the 3-dimensional hypersurface $\Sigma$ are defined as $e^a_\alpha=\partial x^a/\partial y^\alpha$. We define $n^a$ as the unit normal vector on $\Sigma$ pointing in the direction from $\mathcal V^-$ to $\mathcal V^+$. Let $l$ denote the affine parameter (whether proper distance or time) along geodesics perpendicular to $\Sigma$ and set $l$ to be zero at $\Sigma$, negative in the region $\mathcal V^-$, and positive in the region $\mathcal V^+$. The displacement from $\Sigma$ along the geodesics parametrized by $l$ is $dx^a=n^adl$, and $n_a=\epsilon \partial_a l$, where $\epsilon$ is either $1$ or $-1$ when $n^a$ is a spacelike or timelike vector, respectively, i.e., $n^an_a=\epsilon$. 

We shall work with the formalism of tensorial distributions. For any quantity $X$, we define $X=X^+\Theta\left(l\right)+X^-\Theta\left(-l\right)$, where the indexes $\pm$ indicate that the quantity $X^\pm$ is the value of  $X$ in the region $\mathcal V^\pm$, and $\Theta\left(l\right)$ is the Heaviside distribution function, with $\delta\left(l\right)=\partial_l\Theta\left(l\right)$ the Dirac-delta distribution function. Finally, we define $\left[X\right]=X^+|_\Sigma-X^-|_\Sigma$ as the jump of $X$ across $\Sigma$. By definition, this implies that $\left[n^a\right]=\left[e^a_\alpha\right]=0$.

%%%%%%%%%%%%%%%%%%%%%%%%%%%%%%%%%%%%%%%%%%%%%%%%%%%%%%%%%%%%%%%%%%%%%%%%%%%%
\subsection{Geometrical representation}\label{sec:jcgeo}
%%%%%%%%%%%%%%%%%%%%%%%%%%%%%%%%%%%%%%%%%%%%%%%%%%%%%%%%%%%%%%%%%%%%%%%%%%%%

Let us now use the distributional formalism to derive the junction conditions of  Palatini $f(\mathcal R,T)$ gravity in the geometric representation of the theory. We start by defining a metric in both sides of $\Sigma$ as
\begin{equation}\label{geo_def_metric}
g_{ab}=g_{ab}^+\Theta\left(l\right)+g_{ab}^-\Theta\left(-l\right).
\end{equation}
This form of the metric must be used to construct all geometric quantities from this point onwards. In particular, the Christoffel symbols must be obtained from the partial derivatives of $g_{ab}$. This way, taking a partial derivative of Eq. \eqref{geo_def_metric}, one obtains $\partial_cg_{ab}=\partial_cg_{ab}^+\Theta\left(l\right)+\partial_cg_{ab}^-\Theta\left(-l\right)+\epsilon\left[g_{ab}\right]n_c\delta\left(l\right)$. The presence of a term proportional to $\delta\left(l\right)$ raises problems, as it leads to the appearance of products of the form $\Theta\left(l\right)\delta\left(l\right)$ in the Christoffel symbols, and consequently terms proportional to $\delta^2\left(l\right)$ in the Riemann tensor, the latter being singular in the distribution formalism. To prevent these problematic terms from appearing, one must impose the continuity of the metric across $\Sigma$, i.e., $\left[g_{ab}\right]=0$. Defining the induced metric on the hypersurface $\Sigma$ as $h_{\alpha\beta}=g_{ab}e^a_\alpha e^b_{\beta}$, the induced metric on both sides of $\Sigma$ is therefore $h_{\alpha\beta}^+=g_{ab}^+e^a_\alpha e^b_{\beta}$ from $\mathcal V^+$ and $h_{\alpha\beta}^-=g_{ab}^-e^a_\alpha e^b_{\beta}$ from $\mathcal V^-$. Since $\left[g_{ab}\right]=0$, then it follows directly that $h_{ab}$ must be continuous across $\Sigma$, i.e., the first junction condition reads
\begin{equation}\label{geo_jc1}
\left[h_{\alpha\beta}\right]=0 \ .
\end{equation}
This junction condition is the same as in GR, and moreover it also appears in many modified theories of gravity. Taking Eq. \eqref{geo_jc1} into consideration, the partial derivative of Eq. \eqref{geo_def_metric} reduces to
\begin{equation}\label{geo_def_dmetric}
\partial_cg_{ab}=\partial_cg_{ab}^+\Theta\left(l\right)+\partial_cg_{ab}^-\Theta\left(-l\right).
\end{equation}
This result allows one to construct the Christoffel symbols $\Gamma^c_{ab}$ associated to the metric $g_{ab}$ without producing $\delta\left(l\right)$ terms. From these Christoffel symbols, one can then compute the Riemann tensor $R^a_{bcd}$, the Ricci tensor $R_{ab}=R^c_{acb}$, and finally the Ricci scalar $R=g^{ab}R_{ab}$. The Ricci tensor $R_{ab}$ and the Ricci scalar $R$ can then be written in the distributional formalism as
\begin{eqnarray}
R_{ab}&=&R_{ab}^+\Theta\left(l\right)+R_{ab}^-\Theta\left(-l\right)-\nonumber \\
&-&\left(\epsilon e^\alpha_ae^\beta_b\left[K_{\alpha\beta}\right]+n_an_b\left[K\right]\right)\delta\left(l\right),\label{geo_def_Rab} \\
R&=&R^+\Theta\left(l\right)+R^-\Theta\left(-l\right)-2\epsilon\left[K\right]\delta\left(l\right), \label{geo_def_R}
\end{eqnarray}
where $K_{\alpha\beta}=e^a_\alpha e^b_\beta\nabla_a n_b$ is the extrinsic curvature of the hypersurface $\Sigma$, and $K=K^\alpha_\alpha$ is the trace of $K_{\alpha\beta}$. Following Eq. \eqref{def_riccirel}, one can anticipate that the Palatini Ricci tensor $\mathcal R_{ab}$ and the Palatini Ricci scalar $\mathcal R$ will also feature terms proportional to $\delta\left(l\right)$ in the distribution formalism, i.e., it is natural to write
\begin{eqnarray}
\mathcal R_{ab}&=&\mathcal R_{ab}^+\Theta\left(l\right)+\mathcal R_{ab}^-\Theta\left(-l\right)+Z_{ab}\delta\left(l\right), \label{geo_def_PRab} \\
\mathcal R&=&\mathcal R^+\Theta\left(l\right)+\mathcal R^-\Theta\left(-l\right)+Z\delta\left(l\right), \label{geo_def_PR}
\end{eqnarray}
for some tensor $Z_{ab}$ with trace $Z=Z^a_a$, to be computed later. Similarly, we write the stress-energy tensor for the matter fields as $T_{ab}$ with its trace $T$ in the distributional formalism  (note that it is not necessary to do the same with $\Theta_{ab}$ as it can be written in terms of $T_{ab}$ and regular quantities) as
\begin{eqnarray}
T_{ab}&=&T_{ab}^+\Theta\left(l\right)+T_{ab}^-\Theta\left(-l\right)+S_{ab}\delta\left(l\right), \label{geo_def_Tab} \\
T&=&T^+\Theta\left(l\right)+T^-\Theta\left(-l\right)+S\delta\left(l\right), \label{geo_def_T}
\end{eqnarray}
where $S_{ab}$ represents the stress-energy tensor of a potential thin-shell of matter at the separation hypersurface $\Sigma$, with $S=S^a_a$ its trace. Having constructed all relevant quantities in the distributional formalism, we are now ready to continue deducing the junction conditions of the theory.

\subsection{Application to $f(\mathcal{R},T)$ theory}\label{sec:geomjc}

Since the function $f(\mathcal R,T)$ features arbitrary dependences on $\mathcal R$ and $T$, which could in principle be represented in the form of power-laws of $\mathcal R$, $T$ (or even in products between the two variables), the presence of a term proportional to $\delta\left(l\right)$ in Eqs. \eqref{geo_def_PR} and \eqref{geo_def_T} implies that singular products $\delta\left(l\right)^2$ will appear in the function $f(\mathcal R,T)$. To prevent these problematic terms from arising, it is necessary that the quantities $Z$ and $S$ vanish, i.e.,
\begin{equation}\label{geo_jc2_aux}
Z=0, \qquad S=0 \ ,
\end{equation}
for whatever forms the quantities $Z$ and $S$ assume in terms of geometrical quantities. Eqs. \eqref{geo_def_PR} and \eqref{geo_def_T} then reduce to
\begin{eqnarray}
\mathcal R&=&\mathcal R^+\Theta\left(l\right)+\mathcal R^-\Theta\left(-l\right), \label{geo_def_PR_reg} \\
T&=&T^+\Theta\left(l\right)+T^-\Theta\left(-l\right) \ . \label{geo_def_T_reg}
\end{eqnarray}
Furthermore, derivatives of the function $f(\mathcal R,T)$ also appear in Eq. \eqref{def_riccirel}. These derivatives can be expanded via the chain rule into derivatives of both $\mathcal R$ and $T$, as
\begin{equation}\label{def_df}
\partial_a f_\mathcal R=f_{\mathcal R\mathcal R}\partial_a\mathcal R+f_{\mathcal RT}\partial_a T \ ,
\end{equation}
and
\begin{eqnarray}\label{def_ddf}
&&\nabla_a\nabla_b f_\mathcal R= f_{\mathcal R\mathcal R}\nabla_a\nabla_b \mathcal R+f_{\mathcal RT}\nabla_a\nabla_b T+ \\
&&+f_{\mathcal R\mathcal R\mathcal R}\nabla_a\mathcal R\nabla_b\mathcal R+f_{\mathcal RTT}\nabla_aT\nabla_bT+2f_{\mathcal R\mathcal RT}\nabla_{(a}\mathcal R\nabla_{b)}T \ , \nonumber
\end{eqnarray}
where the parenthesis denote index symmetrization, i.e., $X_{(ab)}=\frac{1}{2}\left(X_{ab}+X_{ba}\right)$, for some quantity $X_{ab}$. Taking the partial derivatives of Eqs. \eqref{geo_def_PR_reg} and \eqref{geo_def_T_reg} yields
\begin{eqnarray}
\partial_c \mathcal R&=&\partial_c \mathcal R^+\Theta\left(l\right)+\partial_c\mathcal R^-\Theta\left(-l\right)+\epsilon\left[\mathcal R\right]n_c\delta\left(l\right), \label{geo_def_dPR} \\
\partial_c T&=&\partial_c T^+\Theta\left(l\right)+\partial_cT^-\Theta\left(-l\right)+\epsilon\left[T\right]n_c\delta\left(l\right). \label{geo_def_dT}
\end{eqnarray}
Inserting the results of Eqs. \eqref{geo_def_dPR} and \eqref{geo_def_dT} into Eq. \eqref{def_ddf}, one verifies that, due to the presence of products between $\nabla_a \mathcal{R}$ and $\nabla_a T$ or both, singular terms $\delta\left(l\right)^2$ will appear in Eq. \eqref{def_ddf}. The same problematic terms would also appear in the products $\partial_a f_\mathcal R\partial_b f_\mathcal R$ in Eq. \eqref{def_riccirel}. To avoid these problematic terms, the singular parts of Eqs. \eqref{geo_def_dPR} and \eqref{geo_def_dT} must vanish, i.e., we obtain two more junction conditions of the form
\begin{eqnarray}
\left[\mathcal R\right]&=&0 \ , \label{geo_jc2} \\
\left[T\right]&=&0 \ . \label{geo_jc3}
\end{eqnarray}
The results of Eqs. \eqref{geo_jc2} and \eqref{geo_jc3} allow one to simplify Eqs. \eqref{geo_def_dPR} and \eqref{geo_def_dT} into their respective regular forms, given by
\begin{eqnarray}
\partial_c \mathcal R&=&\partial_c \mathcal R^+\Theta\left(l\right)+\partial_c\mathcal R^-\Theta\left(-l\right), \label{geo_def_dPR_reg} \\
\partial_c T&=&\partial_c T^+\Theta\left(l\right)+\partial_cT^-\Theta\left(-l\right) \ . \label{geo_def_dT_reg}
\end{eqnarray} 
One can now compute the second-order covariant derivatives of $\mathcal R$ and $T$ present in Eq. \eqref{def_ddf} by applying a covariant derivative to Eqs. \eqref{geo_def_dPR_reg} and \eqref{geo_def_dT_reg}, which yields the result
\begin{eqnarray}
\nabla_a\nabla_b\mathcal R&=&\nabla_a\nabla_b\mathcal R^+\Theta\left(l\right)+\nabla_a\nabla_b\mathcal R^-\Theta\left(-l\right) \nonumber  \\
&+&\epsilon\left[\nabla_b \mathcal R\right]n_a\delta\left(l\right) \ , \label{geo_def_ddPR}
\end{eqnarray}
and
\begin{eqnarray}
\nabla_a\nabla_bT &=&\nabla_a\nabla_bT^+\Theta\left(l\right)+\nabla_a\nabla_bT^-\Theta\left(-l\right) \nonumber \\
&+&\epsilon\left[\nabla_b T\right]n_a\delta\left(l\right) \ . \label{geo_def_ddT}
\end{eqnarray}

We now have all the necessary tools to analyze the distributional version of the equations of motion of the theory, starting from Eq. \eqref{geo_field} and using the relation between $\mathcal R_{ab}$ and $R_{ab}$ in Eq. \eqref{def_riccirel}. In particular, the explicit form for the singular part of $\mathcal R_{ab}$, i.e., $Z_{ab}$, can now be obtained from Eq. \eqref{def_riccirel} by keeping only the singular terms, i.e., the terms proportional to $\delta\left(l\right)$, which gives the result
\begin{eqnarray}
&&Z_{ab}=-\epsilon[K_{ab}]-n_an_b[K]-\frac{1}{f_\mathcal R}\Big[f_{\mathcal R\mathcal R}\Big(n_a[\nabla_b\mathcal R]  \label{geo_jc4_aux} \\
&&+\frac{1}{2}g_{ab}n^c[\nabla_c\mathcal R]\Big)+f_{\mathcal RT}\Big(n_a[\nabla_bT]+\frac{1}{2}g_{ab}n^c[\nabla_cT]\Big)\Big] \ .  \nonumber
\end{eqnarray}
Taking the trace of Eq. \eqref{geo_jc4_aux} and using the fact that $Z=0$ from Eq. \eqref{geo_jc2_aux}, one obtains the fourth junction condition as
\begin{equation}\label{geo_jc4}
2\epsilon\left[K\right]+\frac{3}{f_\mathcal R}n^a\left(f_{\mathcal R\mathcal R}\left[\nabla_a\mathcal R\right]+f_{\mathcal RT}\left[\nabla_aT\right]\right)=0 \ .
\end{equation}
Finally, taking the modified field equations in Eq. \eqref{geo_field}, projecting onto the hypersurface $\Sigma$ with the projection vectors (or, equivalently, with the induced metric), and keeping only the singular terms one obtains the fifth and last junction condition under the form
\begin{equation}\label{geo_jc5}
-\epsilon\left[K_{\alpha\beta}\right]+\frac{1}{3}\epsilon\left[K\right]h_{\alpha\beta}=\frac{8\pi+f_T}{f_\mathcal R}S_{\alpha\beta} \ .
\end{equation}
Note that taking the trace of Eq. \eqref{geo_jc5} one recovers $S=0$, a result which was previously anticipated in Eq. \eqref{geo_jc2_aux}.

To summarize, the complete system of junction conditions, in the geometrical representation of the theory, for a general matching between two space-times $\mathcal V^+$ and $\mathcal V^-$ at an hypersurface $\Sigma$ with the possibility of a thin-shell, is composed by a total of five conditions, namely
\begin{eqnarray}
&\left[h_{\alpha\beta}\right]=0, \label{geo_fullset_shell1}  \\
&\left[\mathcal R\right]=0, \label{geo_fullset_shell2}  \\
&\left[T\right]=0,\label{geo_fullset_shell3} \\
&2\epsilon\left[K\right]+\frac{3}{f_\mathcal R}n^a\left(f_{\mathcal R\mathcal R}\left[\nabla_a\mathcal R\right]+f_{\mathcal RT}\left[\nabla_aT\right]\right)=0, \label{geo_fullset_shell4}  \\
&-\epsilon\left[K_{\alpha\beta}\right]+\frac{1}{3}\epsilon\left[K\right]h_{\alpha\beta}=\frac{8\pi+f_T}{f_\mathcal R}S_{\alpha\beta}.\label{geo_fullset_shell5}  \label{eq:junction:v}
\end{eqnarray}
Similarly to what happens in $f(R)$ gravity, the main qualitative difference in the system of junction conditions between the metric and the Palatini approaches to $f(R,T)$ gravity is the absence of a junction condition on $\left[K\right]$ in the Palatini formalism. The relaxation of this restriction allows one to extend the applicability range of the theory. In particular, systems like thin-shell wormholes feature a discontinuous $K$ and are only achievable if the condition $\left[K\right]=0$ is absent.

%%%%%%%%%%%%%%%%%%%%%%%%%%%%%%%%%%%%%%%%%%%%%%%%%%%%%%%%%%%%%%%%%%%%%%%%%%%%
\subsection{Scalar-tensor representation}\label{sec:jcgeo}
%%%%%%%%%%%%%%%%%%%%%%%%%%%%%%%%%%%%%%%%%%%%%%%%%%%%%%%%%%%%%%%%%%%%%%%%%%%%

Let us now repeat the reasoning followed in the previous section but now in the scalar-tensor representation of the Palatini $f(\mathcal R,T)$ gravity theory. The starting point is the same, i.e., we define a metric for the whole spacetime $\mathcal V$ in the distribution formalism as given by Eq. (\ref{geo_def_metric}), and following similar considerations we arrive at the first junction condition in Eq. (\ref{geo_jc1}) as well as to Eqs. (\ref{geo_def_dmetric}), (\ref{geo_def_Rab}), (\ref{geo_def_R}), (\ref{geo_def_Tab}) and (\ref{geo_def_T}). Next, we need to write the two scalar fields $\varphi$ and $\psi$ in the distribution formalism, which take the forms
\begin{eqnarray}
\varphi&=&\varphi^+\Theta\left(l\right)+\varphi^-\Theta\left(-l\right) \ , \label{sca_def_phi} \\
\psi&=&\psi^+\Theta\left(l\right)+\psi^-\Theta\left(-l\right) \ . \label{sca_def_psi}
\end{eqnarray}
Since the potential $V(\varphi,\psi)$ is a function of the scalar fields $\varphi$ and $\psi$ only, the forms of the scalar fields in Eqs. \eqref{sca_def_phi} and \eqref{sca_def_psi} without any proportionality to $\delta\left(l\right)$ guarantee that the potential $V$ is regular and does not feature any singular products of distribution functions. The same regularity is guaranteed to any partial derivative of $V$, following the same argument. Consequently, from Eqs. \eqref{st_eompsi} and \eqref{st_eomphi2}, one verifies that the only possible singular terms arising in these equations are proportional to $S\delta\left(l\right)$. In the absence of other divergent terms to compensate, one is forced to conclude that $S=0$, which was already obtained as the second condition in Eq. (\ref{geo_jc2_aux}). 

As for the first partial derivatives of the scalar fields in the distribution formalism, from Eqs. \eqref{sca_def_phi} and \eqref{sca_def_psi}, we get
\begin{eqnarray}
\partial_c\varphi&=&\partial_c\varphi^+\Theta\left(l\right)+\partial_c\varphi^-\Theta\left(-l\right)+\epsilon\left[\varphi\right]n_c\delta\left(l\right) \ , \label{sca_def_dphi}  \\
\partial_c\psi&=&\partial_c\psi^+\Theta\left(l\right)+\partial_c\psi^-\Theta\left(-l\right)+\epsilon\left[\psi\right]n_c\delta\left(l\right) \ . \label{sca_def_dpsi}
\end{eqnarray}
From the field equations in Eq. \eqref{st_field}, one verifies that the kinetic term associated with the field $\varphi$ features products of the form $\partial_a\varphi\partial_b\varphi$. Since $\partial_a\varphi$ in Eq. \eqref{sca_def_dphi} presents a term proportional to $\delta\left(l\right)$, these products would lead to the appearance of singular terms $\delta\left(l\right)^2$ in the field equations, and thus it is clear that the scalar field $\varphi$ must be continuous to avoid this problem, i.e., $\left[\varphi\right]=0$. However, since the scalar field $\psi$ does not have a kinetic term, there is no obvious reason from the field equations that would force the scalar field $\psi$ to be continuous. Although this would be true for a general scalar-tensor theory defined by an action of the form of Eq. \eqref{st_action}, the argument does not hold when this scalar-tensor theory is defined as an equivalent representation of the Palatini $f(\mathcal R,T)$ gravity. Recall from Eq. \eqref{st_matrixeq} that the scalar-tensor theory is only well defined when the determinant of the matrix $\mathcal M$ is non-zero, for which the definitions of $\mathcal{R}$ and $T$ in terms of the scalar fields $\varphi$ and $\psi$ are invertible, i.e., one must be able to write the scalar fields $\varphi$ and $\psi$ as $\varphi\left(\mathcal R,T\right)$ and $\psi\left(\mathcal R,T\right)$. Now, taking the first and second order derivatives of $\varphi$ that appear in the field equations, one obtains
\begin{equation}\label{sca_def_dphiRT}
\partial_a \varphi=\varphi_{\mathcal R}\partial_a\mathcal R+\varphi_{T}\partial_a T \ ,
\end{equation}
and
\begin{eqnarray}\label{sca_def_ddphiRT}
&&\nabla_a\nabla_b \varphi= \varphi_{\mathcal R}\nabla_a\nabla_b\mathcal R+\varphi_{T}\nabla_a\nabla_b T+ \\
&&+\varphi_{\mathcal R\mathcal R}\nabla_a\mathcal R\nabla_b\mathcal R+\varphi_{TT}\nabla_aT\nabla_bT+2\varphi_{\mathcal RT}\nabla_{(a}\mathcal R\nabla_{b)}T \ , \nonumber
\end{eqnarray}
where the subscripts $\mathcal{R}$ and $T$ denote partial derivatives of $\varphi$ with respect to these variables, respectively. Now, due to the presence of the terms proportional to $\delta\left(l\right)$  in the partial derivatives of $\mathcal{R}$ and $T$ [which were already computed in Eqs. (\ref{geo_def_dPR}) and (\ref{geo_def_dT})],  the products between $\nabla_a \mathcal{R}$, $\nabla_a T$, or both, in Eq. \eqref{sca_def_ddphiRT} will lead to singular terms $\delta\left(l\right)^2$, which must be avoided by forcing the continuity of $\mathcal R$ and $T$, i.e., $\left[\mathcal R\right]=0$ and $\left[T\right]=0$. Since both scalar fields $\varphi$ and $\psi$ are well behaved functions of $\mathcal R$ and $T$ in this equivalent scalar-tensor representation of Palatini $f(\mathcal R,T)$ gravity, these conditions imply the second and third junction conditions in this representation as
\begin{eqnarray}
\left[\varphi\right]&=&0 \ , \label{sca_jc2} \\
\left[\psi\right]&=&0 \ . \label{sca_jc3}
\end{eqnarray}
Junction conditions of this form are expected in any scalar-tensor theory of gravity for which the action features a kinetic term for the scalar fields. Even though only the scalar field $\varphi$ features a kinetic term, the junction condition is true for $\psi$ as well to preserve the equivalence between the geometrical and the scalar tensor representations of the theory. The first and second order derivatives of $\varphi$ then become
\begin{eqnarray}
\partial_c\varphi&=&\partial_c\varphi^+\Theta\left(l\right)+\partial_c\varphi^-\Theta\left(-l\right), \label{sca_def_dphi_reg} \\
\nabla_a\nabla_b\varphi&=&\nabla_a\nabla_b\varphi^+\Theta\left(l\right)+\nabla_a\nabla_b\varphi^-\Theta\left(-l\right) \nonumber \\
&+&\epsilon\left[\nabla_b\varphi\right]n_a\delta\left(l\right) \ , \label{sca_def_ddphi}
\end{eqnarray}
respectively.

We have now obtained all the necessary quantities to analyze the field equations in Eq. \eqref{st_field} and extract the remaining junction conditions. Taking a projection of the field equations into the hypersurface $\Sigma$ using the projection vectors (or, equivalently, the induced metric), and keeping only the terms proportional to $\delta\left(l\right)$ one obtains
\begin{equation}\label{sca_jc4_aux}
\varphi\left(-\epsilon\left[K_{\alpha\beta}\right]+h_{\alpha\beta}\epsilon\left[K\right]\right)+h_{\alpha\beta}n^c\left[\nabla_c\varphi\right]=\left(8\pi+\psi\right)S_{\alpha\beta} \ .
\end{equation}
The field equation in Eq. \eqref{sca_jc4_aux} does not constitute a junction condition \textit{per se}, but includes two separate junction conditions. The first of these conditions is obtained by taking the trace of Eq. \eqref{sca_jc4_aux} and using the condition $S=0$, which yields the fourth junction condition as
\begin{equation}\label{sca_jc4}
2\epsilon\left[K\right]+\frac{3}{\varphi}n^c\left[\nabla_c\varphi\right]=0 \ .
\end{equation}
Finally, the result of Eq. \eqref{sca_jc4} can be used to eliminate $\left[\nabla_c\varphi\right]$ from Eq. \eqref{sca_jc4_aux} in terms of $\left[K\right]$, from which one obtains the fifth and last junction condition as
\begin{equation}\label{sca_jc5}
-\epsilon\left[K_{\alpha\beta}\right]+\frac{1}{3}\epsilon\left[K\right]h_{\alpha\beta}=\frac{8\pi+\psi}{\varphi}S_{\alpha\beta} \ .
\end{equation}
The consistency between Eq. \eqref{sca_jc5} and the condition $S=0$ can be verified by taking the trace of the former to obtain the latter, and thus these two equations represent the same junction condition.

To summarize, the complete set of junction conditions for the scalar-tensor representation of Palatini $f(\mathcal R,T)$ gravity between two spacetimes $\mathcal V^+$ and $\mathcal V^-$ with a possible thin-shell at the separation hypersurface $\Sigma$ is composed of a total of five equations, which are
\begin{eqnarray}
&\left[h_{\alpha\beta}\right]=0, \label{sca_fullset_shell1}\\
&\left[\varphi\right]=0, \label{sca_fullset_shell2}  \\
&\left[\psi\right]=0, \label{sca_fullset_shell3}\\
&2\epsilon\left[K\right]+\frac{3}{\varphi}n^c\left[\nabla_c\varphi\right]=0, \label{sca_fullset_shell4}  \\
&-\epsilon\left[K_{\alpha\beta}\right]+\frac{1}{3}\epsilon\left[K\right]h_{\alpha\beta}=\frac{8\pi+\psi}{\varphi}S_{\alpha\beta}. \label{sca_fullset_shell5}
\end{eqnarray}
It is worth pointing out that this set of junction conditions can be obtained directly from those appearing  in the geometrical representation of the theory obtained in the previous section [i.e., Eqs. (\ref{geo_fullset_shell1}) to (\ref{geo_fullset_shell5})], via the introduction of the definitions of the scalar fields in Eq. \eqref{def_scalars}, which emphasizes the equivalence between the two representations of the theory and gives further support to the coherence of our analysis. 

%%%%%%%%%%%%%%%%%%%%%%%%%%%%%%%%%%%%%%%%%%%%%%%%%%%%%%%%%%%%%%%%%%%%%%%%%%%%
\subsection{Exceptional cases}\label{sec:exc}
%%%%%%%%%%%%%%%%%%%%%%%%%%%%%%%%%%%%%%%%%%%%%%%%%%%%%%%%%%%%%%%%%%%%%%%%%%%%

The sets of junction conditions obtained previously in the systems of Eqs. \eqref{geo_fullset_shell1} to \eqref{geo_fullset_shell5} and \eqref{sca_fullset_shell1} to \eqref{sca_fullset_shell5} for the geometrical and the scalar-tensor representation, respectively, were derived assuming an arbitrary function $f(\mathcal R,T)$ for which all the partial derivatives are non-vanishing to any order. There are, however, particular choices of the function $f(\mathcal R,T)$ for which some of the junction conditions derived previously can be discarded. In this section, we clarify these particular choices.

One of the junction conditions that can be discarded for some particular cases of the function $f(\mathcal{R},T)$ is Eq. \eqref{geo_fullset_shell2}, i.e., $\left[\mathcal R\right]=0$. In Sec. \ref{sec:geomjc}, this junction condition was imposed to avoid the presence of $\delta\left(l\right)^2$ singular terms in the products $\nabla_a \mathcal{R}\nabla_b \mathcal{R}$ arising in the differential terms of Eq. \eqref{def_riccirel}. However, there is an alternative way of avoiding the presence of these terms, which consists of imposing the condition $f_{\mathcal R\mathcal R}=0$ on the function $f(\mathcal R,T)$. With this condition, the terms proportional to $\nabla_a\mathcal R$ disappear from Eqs. \eqref{def_df} and \eqref{def_ddf}, and consequently from Eq. \eqref{def_riccirel}, without the necessity to require $\left[\mathcal R\right]=0$. Similarly, the junction condition $\left[T\right]=0$ was imposed to avoid the presence of the singular terms $\delta\left(l\right)^2$ in the products $\nabla_aT\nabla_bT$ appearing also in Eq. \eqref{def_riccirel}. An alternative way of removing these problematic products is to impose the constraint $f_{\mathcal R T}=0$ on the function $f(\mathcal R,T)$, without the necessity to require $\left[T\right]=0$. Furthermore, if the function $f(\mathcal R,T)$ satisfies both conditions simultaneously, i.e., $f_{\mathcal R\mathcal R}=f_{\mathcal R T}=0$, both junction conditions $\left[\mathcal R\right]=0$ and $\left[T\right]=0$ can be discarded from the system. 

In other theories of gravity, see e.g. \cite{Rosa:2021yym,Rosa:2021teg,Senovilla:2013vra}, the choice of particular cases for which some junction conditions can be discarded results in the appearance of extra terms in the modified field equations that can be interpreted as extra matter components, namely external stresses, momentum fluxes, and the so-called double gravitational layer. Indeed, if $\left[\mathcal R\right]\neq 0$ and $\left[T\right]\neq0$, the last terms on the right-hand side of Eqs. \eqref{geo_def_dPR} and \eqref{geo_def_dT} do not vanish, and the second-order covariant derivatives of these variables will feature extra terms. However, as can be seen from Eq. \eqref{def_ddf}, the second-order derivatives of $\mathcal R$ and $T$ appear multiplied by a factor of $f_{\mathcal R\mathcal R}$ and $f_{\mathcal R T}$, respectively, which are zero for the particular cases of the function $f(\mathcal R,T)$ that allow to discard these junction conditions. Consequently, no extra terms appear in the field equations of Palatini $f(\mathcal R,T)$ gravity, and the final set of junction conditions can be obtained simply as the limit of the general set.

\subsubsection{Case 1: $\left[\mathcal R\right]\neq 0$}

The most general form of the function $f(\mathcal R,T)$ that satisfies the condition $f_{\mathcal R\mathcal R}=0$, and hence allows for $\left[\mathcal R\right]\neq 0$, is given by
\begin{equation}\label{case1f}
f(\mathcal R,T)=\mathcal R\left[\alpha+g(T)\right]-2\Lambda+h(T),
\end{equation}
where $g(T)$ and $h(T)$ are arbitrary functions of $T$, and the parameters $\alpha$ and $\Lambda$ are constants, the latter conveniently chosen to play the role of a cosmological constant. Under this choice of the function $f(\mathcal{R},T)$, the set of junction conditions from Eqs. \eqref{geo_fullset_shell1}, \eqref{geo_fullset_shell3}, \eqref{geo_fullset_shell4}, \eqref{geo_fullset_shell5}  becomes
\begin{eqnarray}
&\left[h_{\alpha\beta}\right]=0, \\
&\left[T\right]=0, \\
&2\epsilon\left[K\right]+\frac{3g'\left(T\right)}{\alpha+g\left(T\right)}n^a\left[\nabla_aT\right]=0, \\
&-\epsilon\left[K_{\alpha\beta}\right]+\frac{1}{3}\epsilon\left[K\right]h_{\alpha\beta}=\frac{8\pi+\mathcal R g'\left(T\right)+h'\left(T\right)}{\alpha+g\left(T\right)}S_{\alpha\beta}. \label{eq:junction:v}
\end{eqnarray}
where we use primes to denote derivatives of each function with respect to its argument. The form of the function $f(\mathcal R,T)$ given in Eq. \eqref{case1f} is general enough for one to define a scalar-tensor representation. Indeed, the requirement $f_{\mathcal R\mathcal R}f_{TT}\neq f_{\mathcal R T}^2$ imposes the constraint $g'\left(T\right)\neq 0$. Thus, if the function $g(T)$ is at least linear in $T$, a scalar-tensor representation is well defined. In this representation, since $\varphi=f_\mathcal R$ and $f_{\mathcal R\mathcal R}=0$, one obtains $\varphi_\mathcal R=0$, i.e., the scalar field $\varphi$ depends solely in $T$. Since $\left[T\right]=0$ is still a requirement in this particular case, no junction conditions are discarded in the scalar-tensor representation.

\subsubsection{Case 2: $\left[T\right]\neq 0$}

The most general form of the function $f(\mathcal R,T)$ satisfying the condition $f_{\mathcal R T}=0$, and thus allowing for $\left[T\right]\neq 0$, takes the following separable form
\begin{equation}\label{case2f}
f(\mathcal R,T)=f(\mathcal R)+g(T),
\end{equation}
Note that both functions $f(\mathcal{R})$ and $g(T)$ are arbitrary, as the only necessary requirement to fulfill the condition $f_{\mathcal RT}=0$ is the absence of crossed terms. The full set of junction conditions becomes for this case
\begin{eqnarray}
&\left[h_{\alpha\beta}\right]=0, \\
&\left[\mathcal R\right]=0, \\
&2\epsilon\left[K\right]+3\frac{f''\left(R\right)}{f'\left(R\right)}n^a\left[\nabla_a\mathcal R\right]=0, \\
&-\epsilon\left[K_{\alpha\beta}\right]+\frac{1}{3}\epsilon\left[K\right]h_{\alpha\beta}=\frac{8\pi+g'\left(T\right)}{f'\left(R\right)}S_{\alpha\beta}. 
\end{eqnarray}
Similarly to the previous case, one is also able to obtain an equivalent scalar-tensor representation associated to the function $f(\mathcal R,T)$ in Eq. \eqref{case2f}. The requirement $f_{\mathcal R\mathcal R}f_{TT}\neq f_{\mathcal R T}^2$ in this case provides the condition $f''(\mathcal R)g''(T)\neq0$, i.e., both the functions $f(\mathcal R)$ and $g(T)$ must be at least quadratic in $\mathcal R$ and $T$, respectively. Furthermore, since $\varphi=f_\mathcal R$ and $f_{\mathcal R T}=0$, one obtains that $\varphi_T=0$, and the scalar field $\varphi$ depends solely in the quantity $\mathcal R$. Since $\left[\mathcal R\right]=0$ in this case, the condition $\left[\varphi\right]=0$ cannot be discarded in the scalar-tensor representation.

\subsubsection{Case 3: $\left[\mathcal R\right]\neq 0$ and $\left[T\right]\neq 0$}

Finally, the most general form of the function $f(\mathcal R,T)$ that satisfies both the conditions $f_{\mathcal R\mathcal R}=0$ and $f_{\mathcal R T}=0$ simultaneously, thus permitting $\left[\mathcal R\right]\neq 0$ and $\left[T\right]\neq 0$, is the following
\begin{equation}\label{case3f}
f(\mathcal R,T)=\alpha \mathcal R-2\Lambda+g(T).
\end{equation}
Note that in this particular case the conformal factor $f_\mathcal R=\alpha$ is a constant, which implies from Eq. \eqref{def_riccirel} that the Palatini and the metric Ricci scalars $\mathcal R_{ab}$ and $R_{ab}$ coincide, i.e., the connection $\hat\Gamma^c_{ab}$ becomes Levi-Civita to the metric $g_{ab}$. In this case, the full set of junction conditions becomes
\begin{eqnarray}
&\left[h_{\alpha\beta}\right]=0, \\
&\left[K\right]=0, \\
&-\epsilon\left[K_{\alpha\beta}\right]=\frac{8\pi+g'\left(T\right)}{\alpha}S_{\alpha\beta}.
\end{eqnarray}
It is important to remark that for this particular choice of the function $f(\mathcal R,T)$, the trace of the extrinsic curvature must be continuous across $\Sigma$. This result is consistent with the previously mentioned result that $\mathcal R_{ab}$ and $R_{ab}$ coincide, thus corresponding to a metric theory $f(R,T)$, for which one expects $\left[K\right]=0$, see \cite{Rosa:2021teg}. It is also worth noticing that the function $f(\mathcal R,T)$ in Eq. \eqref{case3f} is not general enough to allow for the transformation to a scalar-tensor representation, as the requirement $f_{\mathcal R\mathcal R}f_{TT}\neq f_{\mathcal R T}^2$ is never satisfied if $f_{\mathcal R\mathcal R}$ and $f_{\mathcal R T}$ vanish simultaneously.

%%%%%%%%%%%%%%%%%%%%%%%%%%%%%%%%%%%%%%%%%%%%%%%%%%%%%%%%%%%%%%%%%%%%%%%%%%%%
\section{An application to traversable thin-shell wormholes}\label{sec:apps}
%%%%%%%%%%%%%%%%%%%%%%%%%%%%%%%%%%%%%%%%%%%%%%%%%%%%%%%%%%%%%%%%%%%%%%%%%%%%

The formalism and conditions developed in the previous sections allows one to consider any applications of interest in which the matching of two space-times is needed to model a particular physical system. Prominent among them is the one of traversable thin-shell wormholes. In such a case, one scisses the exterior region to (the same or different) two black hole space-times, i.e.,  above its would-be event horizon, and patches their respective boundaries at a shell, which in the present case is dubbed as the throat. The interest in this kind of construction lies on the restoration of the geodesic completeness of the original space-times, while the prize to be paid is typically a potential violation of the energy conditions at the throat, at least within GR \cite{Lobo:2005us}.

We consider a similar construction as the one of Ref. \cite{Lobo:2020vqh} for Palatini $f(\mathcal{R})$ gravity, where two external pieces of the Schwarzschild space-time are glued at the throat, and the construction is subsequently extended to two Reissner-Nordstr\"om space-times (i.e. charged under Maxwell electrodynamics). The metric on each space-time is thus written in spherical coordinates $\left(t,r_{\pm},\theta,\phi\right)$ as
\begin{equation}
ds^2=-A^{\pm}(r_{\pm})dt^2+\frac{dr_{\pm}^2}{A^{\pm}(r_{\pm})} + r_{\pm}^2d\Omega^2,
\end{equation}
where $A^{\pm}(r_{\pm})=1-2\tfrac{M_{\pm}}{r_{\pm}}+\tfrac{Q_{\pm}^2}{r_{\pm}^2}$ is the metric on each side of the shell, with $\{M_{\pm},Q_{\pm}\}$ the corresponding masses and electric charges, and $d\Omega^2=d\theta^2+\sin^2\theta d\phi^2$ is the line element of the 2-spheres.  Note that one does not necessarily need to impose neither $M_+=M_-$ nor $Q_+=Q_-$, thus allowing for the existence of asymmetric wormholes, i.e., different space-times on each side of the throat \cite{Guerrero:2021pxt}.
Since for any of the space-times above one has $T=0$, the junction condition in Eq. (\ref{geo_fullset_shell3}) is automatically satisfied, while the trace of Eq. \eqref{geo_fullset_shell5} implies the necessity of $S=0$ as well. In the Palatini $f(\mathcal{R})$ case this means that $\kappa^2/f_{\mathcal{R}}\vert_{T=0}=$constant  everywhere, regardless of the specific shape of the $f(\mathcal{R})$ Lagrangian chosen. This conclusion still holds in the present $f(\mathcal{R},T)$ case, since from the Eqs. (\ref{def_tab}) and (\ref{def_theta}) and writing Maxwell Lagrangian as $\mathcal{L}_m=-\tfrac{1}{4} F_{ab}F^{ab}$, one can compute the tensor $\Theta_{ab}$ in the present case as
\begin{equation}
\Theta_{ab}=-T_{ab}={F_a}^c F_{cb}-\frac{1}{4}g_{ab} F_{cd}F^{cd} \ ,
\end{equation}
Now, from the definition in Eq. (\ref{eq:taumunu}) we can see the effect of this result in the behaviour of $\tau=T-\tfrac{f_T}{\kappa^2}(T+\Theta)$ (and hence of curvature  via Eq. (\ref{eq:curtra})): for Schwarzschild space-times, $T_{ab}=0$ so $\tau=0$ and the conclusion above on the constancy of $f_{\mathcal{R}}$ is unmodified, while for Reissner-Nordstr\"om space-times the same conclusion is reached no matter the shape of the $f(\mathcal{R},T)$ Lagrangian, since the tracelessness of $T_{\mu\nu}$ translates into the tracelessness of both $\Theta_{\mu\nu}$ and $\tau_{\mu\nu}$ and hence no new dynamics is introduced.

On the shell $\Sigma$, parametrized by
\begin{equation}
ds_{\Sigma}^2=-d\Upsilon^2+\varrho^2(T)d\Omega^2,
\end{equation}
where $\Upsilon$ and $\varrho$ are the shell's temporal and radial coordinates, respectively, the continuity of the metric along $\Sigma$  does introduce a constraint, $2(M_+-M_-)\varrho=Q_+^2-Q_-^2$. On the other hand, the junction condition in Eq. (\ref{eq:junction:v}) introduces a relation between the metric functions restricted to the shell, and the behaviour of the matter fields there. For this purpose, two extra equations are needed, which correspond to the projection on the shell of the Bianchi identities, namely (see Appendix of Ref. \cite{Senovilla:2013vra}):
\begin{eqnarray}
(K_{\rho\sigma}^+ + K_{\rho\sigma}^-)\mathcal{G}^{\rho\sigma}&=&2n^\rho n^\sigma [\mathcal{R}_{\rho\sigma}]-[\mathcal{R}] \\
D^{\alpha} \mathcal{G}_{\alpha\beta}&=&-n^{\alpha}h_{\beta}^{\sigma}[\mathcal{R}_{\alpha\sigma}],
\end{eqnarray}
where $D_{\rho} \equiv {h_\rho}^{\alpha}\nabla_{\alpha}$ is the covariant derivative in $\Sigma$ and $\mathcal G_{\alpha\beta}=e^a_\alpha e^b_\beta\left(\mathcal R_{ab}-\frac{1}{2}g_{ab}\mathcal R\right)$. Following a similar analysis as in Ref. \cite{Olmo:2020fri} replacing $T_{ab}$ by $\tau_{ab}$ via Eq. (\ref{eq:taumunu}) one finds
\begin{equation} \label{eq:cons}
D^\alpha s_{\alpha\beta}=-n^\alpha h^\sigma_\beta[\tau_{\alpha\sigma}],
\end{equation}
where $s_{\alpha\beta}=e^a_\alpha e^b_\beta \tau_{ab}$ is the projection of the effective stress-energy tensor $\tau_{ab}$ on the shell. In the present case of Schwarzschild/Reissner-Nordstr\"om configurations, the vanishing of $T$ entails that $\tau_{ab}=T_{ab}$ and thus $s_{\alpha\beta}=S_{\alpha\beta}$. The bottom line of the above discussion is that, upon a redefinition of the constant $\kappa^2$ of the theory, traversable wormholes (either symmetric or asymmetric) coincide with those of Palatini $f(\mathcal{R})$ gravity analyzed in \cite{Lobo:2020vqh,Guerrero:2021pxt}.

Things differ abruptly when non-linear electromagnetic fields are considered. In the simplest case in which the Maxwell Lagrangian is extended to a general function $\varphi(X)$ with $X=-\tfrac{1}{2}F_{ab}F^{ab}$ (so that Maxwell corresponds to $\varphi(X)=X/2$), then the trace of the effective stress-energy tensor reads
\begin{equation}
\tau=T+\frac{8f_T}{\kappa^2 }X^2\varphi_{XX}
\end{equation}
where now the fact that $T=4(-X\varphi_X+\varphi) \neq 0$ makes the (non-vanishing) $f_T$-contributions to alter in all cases the relation between the metric functions and the energy density and pressure of the matter fields on the shell via the junction conditions, as compared to the $f(\mathcal{R})$ counterparts. Since the external solutions on both $\mathcal{V}^{\pm}$ can be found under analytical forms for $f(\mathcal{R})$ and $f(\mathcal{R},T)$ gravity, one can find new traversable wormhole solutions in both theories, enhancing the variety of symmetric and asymmetric solutions within this formalism. In particular, there is plenty of room to play with combinations of non-linear electrodynamics on each side of the throat, including transfer of fluxes across of it given the non-conservation of the effective stress-energy tensor, seeking for solutions of this kind which are stable and satisfy canonical energy conditions.  Since this issue goes beyond the scope of this work, we shall implement these ideas in future projects.

\section{Conclusion}\label{sec:concl}

The finding of suitable junction conditions for every (modified) theory of gravity is of interest in order to extract useful modelling of different physical systems, and to seek  deviations with respect to GR expectations. In this work we have analyzed the junction conditions of yet one more theory, namely, Palatini $f(\mathcal{R},T)$ gravity. Working dually in the usual geometrical representation of the theory as well as on its physically equivalent scalar-tensor counterpart, we have found that the modifications as compared to Palatini  $f(\mathcal{R})$ junction conditions consist of a $f_T$-correction term in two of them. At the same time, we further confirm the absence of a junction condition on $[K]$ as compared to the metric version of this theory, which allows a larger flexibility in running specific applications of interest within these theories.

The particular forms of the function $f(\mathcal R,T)$ for which some of the junction conditions can be discarded  from the full set (``exceptional" cases) were also analyzed. In other theories of gravity, see e.g. \cite{Rosa:2021yym,Rosa:2021teg,Senovilla:2013vra}, some of these exceptional cases were shown to give rise to extra terms in the stress-energy tensor associated with external tensions, momentum fluxes, and double gravitational layers. Interestingly, we have shown that the same does not happen in Palatini $f(\mathcal R,T)$ gravity, as the terms potentially contributing to the extra matter components vanish identically in this theory. This allows one to select simpler forms of the action that simplify the set of junction conditions, without the drawback of further complicating the matter sector. 

Among such applications we briefly discussed the one of traversable thin-shell wormholes, which can be implemented both in reflection-symmetric and asymmetric scenarios, and which, similarly as their $f(\mathcal{R})$ cousins, could be of interest from the point of view of both (double) shadows in gravitational light deflection \cite{Wielgus:2020uqz}, and in echoes of gravitational wave emission from binary mergers of horizonless compact objects \cite{Cardoso:2016oxy}. For the simplest case of Maxwell fields, the tracelessness of its stress-energy tensor implies that the formalism yields the same results as their $f(\mathcal{R})$ counterpart, forcing one to upgrade the matter field description up to non-linear electromagnetic fields, where now any non-trivial $f(\mathcal{R},T)$ function will introduce modifications to the (shell) dynamics. 

Another topic of particular interest for astrophysical applications is  the definition of surfaces in stellar bodies within these theories. Such bodies are typically modeled as a fluid-filled interior (where the corresponding TOV equations are solved) and matched to an external Schwarzschild vacuum space-time. Beyond GR this construction may harbor theoretical difficulties, for instance, in those theories having extra dynamical degrees of freedom, like metric  $f(R)$ gravity \cite{AparicioResco:2016xcm} due to the contribution of the scalar field to the total energy of the system, while in Palatini $f(\mathcal{R})$ gravity the difficulty lies instead on the fact that polytropic equations of state may develop local curvature divergences at the stellar surface \cite{Barausse:2007pn}. The latter problem is solved (at least within a range of polytropic indices of physical interest) when considering the junction conditions formalism to correctly perform the matching \cite{Olmo:2020fri}. It would therefore be  necessary to see how such polytropic models in stellar surfaces behave when running the same exercise in the present $f(\mathcal{R},T)$ case. Moreover, a similar analysis could be carried out for other solutions of 
theoretical and cosmological interest with sharp transitions for the inner to outer regions, such as in cosmic strings, domain walls or brane-world scenarios. And finally, the junction conditions derived here could be of interest in order to model the instability of Cauchy horizons in black holes arising in $f(\mathcal{R},T)$ theories, similarly as done in \cite{Capozziello:2009pi,Capozziello:2011gw,Capozziello:2013gza} for other extensions of GR. In the present case, the non-conservation of the stress-energy tensor of the matter fields  allows for interesting new possibilities in all such scenarios, such as the existence of energy fluxes across the matching shell, whose implementation is also worth exploring. \\

%%%%%%%%%%%%%%%%%%%%%%%%%%%%%%%%%%%%%%%%%%%%%%%%%%%%%%%%%%%%%%%%%%%%%%%%%%%%

%%%%%%%%%%%%%%%%%%%%%%%%%%%%%%%%%%%%%%%%%%%%%%%%%%%%%%%%%%%%%%%%%%%%%%%%%%%%
\begin{acknowledgements}
JLR is supported by the European Regional Development Fund and the programme Mobilitas Pluss (MOBJD647), and thanks the Department of Theoretical Physics at the Complutense University of Madrid for their hospitality during the elaboration of this work. DRG is funded by the \emph{Atracci\'on de Talento Investigador} programme of the Comunidad de Madrid (Spain) No. 2018-T1/TIC-10431, and acknowledges further support from the Ministerio de Ciencia, Innovaci\'on y Universidades (Spain) project No. PID2019-108485GB-I00/AEI/10.13039/501100011033 (``PGC Generaci\'on de Conocimiento") and the FCT projects No. PTDC/FIS-PAR/31938/2017 and PTDC/FIS-OUT/29048/2017.  This work is also supported by the project PROMETEO/2020/079 (Generalitat Valenciana), and the Edital 006/2018 PRONEX (FAPESQ-PB/CNPQ, Brazil, Grant 0015/2019). This article is based upon work from COST Action CA18108, supported by COST (European Cooperation in Science and Technology).
\end{acknowledgements}
%%%%%%%%%%%%%%%%%%%%%%%%%%%%%%%%%%%%%%%%%%%%%%%%%%%%%%%%%%%%%%%%%%%%%%%%%%%%

%%%%%%%%%%%%%%%%%%%%%%%%%%%%%%%%%%%%%%%%%%%%%%%%%%%%%%%%%%%%%%%%%%%%%%%%%%%%

\end{document}